# Minkowski domain walls in hyperbolic metamaterials

Igor I. Smolyaninov, Yu-Ju Hung

*Department of Electrical and Computer Engineering, University of Maryland, College Park, MD 20742, USA*

*phone: 301-405-3255; fax: 301-314-9281; e-mail: smoly@umd.edu*

**Minkowski domain walls are being actively considered in gravitation theory. They may form during a vacuum phase transition, or as a result of braneworld collision. Despite having interesting physical properties, Minkowski domain walls had remained in the theoretical domain only since their first introduction a few decades ago. Here we demonstrate how to make an electromagnetic analogue of a Minkowski domain wall using hyperbolic metamaterials. We analyze electromagnetic field behavior at the wall, and present a simple experimental model of "Minkowski domain wall" formation due to "collision" of two Minkowski spaces.**

# 1. Introduction.

Optics of metamaterials provides us with many possibilities to design effective space-times with highly unusual properties [1-6]. Hyperbolic metamaterials are especially interesting in this respect, since propagation of monochromatic extraordinary light in a hyperbolic metamaterial is similar to propagation of massive particles in a three dimensional effective Minkowski spacetime, in which the role of timelike variable is played by one of the spatial coordinates [7-9]. It appears that physics at the boundaries of this effective spacetime is also quite unusual. For example, a planar interface between a hyperbolic metamaterial and a bulk metal may act as an "electromagnetic black hole" [10]. Here we will present another interesting example of unusual physics at the hyperbolic metamaterial interfaces: an electromagnetic "Minkowski domain wall" (Fig.1).

Despite having very interesting physical properties, Minkowski domain walls had remained in the gravitation theory domain only since their first introduction a few decades ago [11]. They may form during a vacuum phase transition [12], or as a result of braneworld collision [13,14]. Thus, there exist quite a few good reasons to try and build an experimental model of a Minkowski domain wall, and study its physical properties. Moreover, recent demonstration that in the presence of very large magnetic field physical vacuum itself behaves as a hyperbolic metamaterial [15] provides an additional incentive.

Let us start with a quick summary of the effective Minkowski spacetime description of light propagation inside hyperbolic metamaterials. This model is described in great detail in refs.[7,8,15]. Let us assume that a uniaxial anisotropic metamaterial has constant dielectric permittivities $\varepsilon_x = \varepsilon_y = \varepsilon_1 > 0$ and $\varepsilon_z = \varepsilon_2 < 0$ in a frequency range around $\omega = \omega_0$ (we also assume this material to be non-magnetic: $\mu = 1$). We will consider the extraordinary ($\vec{E}$ parallel to the plane defined by the k–vector of the wave and the optical axis) component of the field and introduce a "scalar" wave function as $\varphi = E_z$. Since hyperbolic metamaterials exhibit large dispersion, we will work in the frequency domain and write the macroscopic Maxwell equations [16] as

$$\frac{\omega^2}{c^2}\vec{D}_\omega = \vec{\nabla} \times \vec{\nabla} \times \vec{E}_\omega \quad \text{and} \quad \vec{D}_\omega = \vec{\varepsilon}_\omega \vec{E}_\omega \qquad (1)$$

If the metamaterial is illuminated by coherent CW laser field at frequency $\omega_0$, the spatial distribution of the extraordinary field $\varphi_\omega$ at this frequency is described by the following wave equation for $\varphi_\omega$:

$$-\frac{\partial^2 \varphi_\omega}{\varepsilon_1 \partial z^2} + \frac{1}{(-\varepsilon_2)}\left(\frac{\partial^2 \varphi_\omega}{\partial x^2} + \frac{\partial^2 \varphi_\omega}{\partial y^2}\right) = \frac{\omega_0^2}{c^2}\varphi_\omega = \frac{m^{*2} c^2}{\hbar^2}\varphi_\omega \qquad (2)$$

This equation coincides with the 3D Klein-Gordon equation describing a massive scalar field $\varphi_\omega$, in which the spatial coordinate $z$ behaves as a "timelike" variable. Thus, eq.(2) describes world lines of massive particles which propagate in a flat (2+1) Minkowski spacetime. If an interface between two metamaterials (Fig.1) is engineered so that the orientation of the optical axis changes direction across the interface, a "Minkowski domain wall" will be created.

## 2. Results.

Let us analyze in detail the most striking example of an electromagnetic Minkowski domain wall in which the orientation of timelike $z$ variable (the optical axis) is rotated by 90 degrees, as shown in Fig.1. We will assume that the medium remains uniaxial everywhere, and that $\varepsilon_x$, $\varepsilon_y$ and $\varepsilon_z$ change continuously as a function of $z$ through a very thin transition layer around the domain wall located at z=0, as shown in Fig.2. Thus, on both sides of the wall we will have the same wave equation for the extraordinary field $\varphi_\omega$. Taking into account $z$ derivatives of $\varepsilon_1$ and $\varepsilon_2$, eq.(1) results in the following equation for $\varphi_\omega$:

$$-\frac{\partial^2 \varphi_\omega}{\varepsilon_1 \partial z^2} + \frac{1}{(-\varepsilon_2)}\left(\frac{\partial^2 \varphi_\omega}{\partial x^2} + \frac{\partial^2 \varphi_\omega}{\partial y^2}\right) + \left(\frac{1}{\varepsilon_1^2}\left(\frac{\partial \varepsilon_1}{\partial z}\right) - \frac{2}{\varepsilon_1 \varepsilon_2}\left(\frac{\partial \varepsilon_2}{\partial z}\right)\right)\left(\frac{\partial \varphi_\omega}{\partial z}\right) + \\ + \frac{\varphi_\omega}{\varepsilon_1 \varepsilon_2}\left(\frac{1}{\varepsilon_1}\left(\frac{\partial \varepsilon_1}{\partial z}\right)\left(\frac{\partial \varepsilon_2}{\partial z}\right) - \left(\frac{\partial^2 \varepsilon_2}{\partial z^2}\right)\right) = \frac{\omega_0^2}{c^2}\varphi_\omega \qquad (3)$$

Let us consider a plane wave solution in the $xy$ direction with an in-plane wave vector $k$. Now we may introduce a new "wave function" $\psi$ as $\varphi_\omega = \psi \varepsilon_1^{1/2}/\varepsilon_2$ and obtain a new wave equation for $\psi$:

$$-\frac{\partial^2 \psi}{\partial z^2} + \psi\left(\frac{\varepsilon_1 k^2}{\varepsilon_2} - \frac{\varepsilon_1 \omega_0^2}{c^2} - \frac{1}{2\varepsilon_1}\left(\frac{\partial^2 \varepsilon_1}{\partial z^2}\right) + \frac{3}{4\varepsilon_1^2}\left(\frac{\partial \varepsilon_1}{\partial z}\right)^2\right) = 0, \qquad (4)$$

where the second term may be considered as an effective "potential energy". Unless there is a special physical reason, linear behaviour $\varepsilon_1 = -\alpha z$ and $\varepsilon_2 = \beta z$ may be assumed inside the thin transition layer near $z=0$ (similar to [17]) resulting in the effective potential energy behaving as $V \sim 3/4z^2$ near the Minkowski domain wall. Such an infinite repulsive potential barrier cannot be penetrated, resulting in perfect isolation of the two Minkowski half spaces separated by the domain wall. This result is interesting since penetration of the wall would violate the metamaterial version of "causality" [8].

We should also note that similar to the braneworld model [13,14], the electromagnetic Minkowski domain wall may support surface electromagnetic states [18] localized on the wall. Let us consider a Minkowski domain wall shown in Fig.3. The optical axis of the hyperbolic metamaterial is oriented at angles $\alpha_1$ and $\alpha_2$ with respect to the wall in the left and right Minkowski half spaces, respectively. Due to lack of symmetry, this configuration is cumbersome to solve analytically. On the other hand, we may use the dispersion law of extraordinary photons in the hyperbolic metamaterial [19]

$$\frac{k_z^2}{\varepsilon_1} + \frac{k_x^2}{\varepsilon_2} = \frac{\omega^2}{c^2} \qquad (5)$$

(where $\varepsilon_x = \varepsilon_y = \varepsilon_1 > 0$ and $\varepsilon_z = \varepsilon_2 < 0$) to derive the condition for surface wave existence. For a surface wave propagating along the domain wall shown in Fig.3 eq.(5) can be re-written as

$$\frac{(q\cos\alpha_i + q_{i\perp}\sin\alpha_i)^2}{\varepsilon_1} + \frac{(q\sin\alpha_i - q_{i\perp}\cos\alpha_i)^2}{\varepsilon_2} = \frac{\omega^2}{c^2} \qquad (6)$$

where $q$ is the wave vector of the surface wave, $\text{Im}(q_{i\perp}) \neq 0$ on both sides of the domain wall, and $i=1,2$. By solving eq.(6) as a quadratic equation with respect to $q_{i\perp}$, a condition for surface wave to exist may be found as

$$\frac{q^2}{\varepsilon_1(-\varepsilon_2)} + \frac{\omega^2 \sin^2(\alpha_i)}{c^2}\left(\frac{1}{\varepsilon_1} + \frac{1}{(-\varepsilon_2)}\right) < \frac{\omega^2}{(-\varepsilon_2)c^2} \qquad (7)$$

This condition is evidently satisfied for small $q < \varepsilon_1 \omega^2/c^2$ and small $\alpha_1 \neq \alpha_2$. This result is easy to understand. While the wave vector of extraordinary photons is not limited from the top, $k_z$ of a freely propagating extraordinary photon cannot be smaller than $\varepsilon_1 \omega^2/c^2$ (see eq.(5)). Therefore, a domain wall having small $\alpha_1 \neq \alpha_2$ supports surface wave solutions having $q$ below this limiting value. Similar to the "braneworld collision" scenario [13,14], these surface waves form a low-dimensional metamaterial space, which is localized on a Minkowski domain wall formed by the "collision" of two higher dimensional Minkowski spaces.

As demonstrated by our earlier papers [8,10], it is relatively easy to emulate various 3D hyperbolic metamaterial geometries by planar plasmonic metamaterial arrangements. While rigorous description of such metamaterials in terms of Diakonov surface plasmons may be found in ref. [20], qualitative analogy between 3D and 2D metamaterials may be explained as follows. A surface plasmon (SP) propagating over a flat metal-dielectric interface may be described by its well-known dispersion relation

$$k_p = \frac{\omega}{c}\left(\frac{\varepsilon_d \varepsilon_m}{\varepsilon_d + \varepsilon_m}\right)^{1/2} \qquad (8)$$

where metal layer is considered to be thick, and $\varepsilon_m(\omega)$ and $\varepsilon_d(\omega)$ are the frequency-dependent dielectric constants of the metal and dielectric, respectively [18]. Thus, similar to the 3D case, we may introduce an effective 2D dielectric constant $\varepsilon_{2D}$, which characterizes the way in which SPs perceive the dielectric material deposited onto the metal surface. By requiring that $k_p = \varepsilon_{2D}^{1/2}\omega/c$, we obtain

$$\varepsilon_{2D} = \left(\frac{\varepsilon_d \varepsilon_m}{\varepsilon_d + \varepsilon_m}\right) \qquad (9)$$

Equation (9) makes it obvious that depending on the plasmon frequency, SPs perceive the dielectric material bounding the metal surface (for example a PMMA layer) in drastically different ways. At low frequencies $\varepsilon_{2D} \approx \varepsilon_d$, so that plasmons perceive a PMMA layer as a dielectric. On the other hand, at high enough frequencies at which $\varepsilon_d(\omega) > -\varepsilon_m(\omega)$ (this happens around $\lambda_0 \sim 500$ nm for a PMMA layer) $\varepsilon_{2D}$ changes sign and becomes negative. Thus, around $\lambda_0 \sim 500$ nm plasmons perceive a PMMA layer on gold as an "effective metal". As a result, around $\lambda_0 \sim 500$ nm plasmons perceive a PMMA stripe pattern from Fig.4(b) as a layered hyperbolic metamaterial shown in Fig.1(b) (note however that dimensionality of the problem is reduced by one compared to the original theoretical consideration). Fabrication of such plasmonic hyperbolic metamaterials in two dimensions requires only very simple and common lithographic techniques [21].

The "colliding braneworld" scenario can be realized as a simple extension of our earlier experiments simulating the spacetime geometry in the vicinity of big bang [8]. Plasmon rays are launched into the hyperbolic metamaterial near $r=0$ point via the central phase matching structure marked with an arrow in Figs.4(a,b). Similar to the world line behavior near the big bang, plasmonic rays or "world lines" indeed increase their spatial separation as a function of "timelike" radial coordinate $r=\tau$. The point (or moment) $r=\tau=0$ corresponds to a moment of the toy "big bang". In the "colliding braneworld" experiment shown in Figs.4(d,e) similar concentric PMMA ring patterns has been created in order to emulate expanding spacetime. When the two concentric ring patterns touch each other ("collide"), a "Minkowski domain wall" is created, in which the metallic stripes touch each other at a small angle, as shown in Fig.3(b). While these results are preliminary, and more sophisticated experimental verification of this idea using 3D hyperbolic metamaterials would be highly desirable, our experimental results are consistent with the presence of surface states at the Minkowski domain wall. Since surface states localized on the wall are supposed to have low wave vectors, excitation of these states at the "collision point" would lead to increased scattering into regular photons, which is quite obvious from Fig.4(e). We should also mention that further analysis of such experiments requires detailed examination of potential role of the optical Tamm states (see for example recent ref. [22]), which may also exist at the interface.

**3. Conclusion**

In conclusion, we have examined metamaterial optics at the electromagnetic "Minkowski domain wall" formed at the interface between two hyperbolic metamaterials. Theoretical picture of the electromagnetic field behavior at the wall is supported by experimental observations. Our results are somewhat similar to the Minkowski domain wall behaviour in gravitation theory.

**Figure Captions**

**Figure 1.** (a) Schematic view of the electromagnetic "Minkowski domain wall" between two hyperbolic metamaterials: if an interface between two metamaterials is engineered so that the orientation of the optical axis changes direction across the interface, a Minkowski domain wall is created. (b) Experimental geometry, which realizes the "Minkowski domain wall".

**Figure 2.** Coordinate behavior of the diagonal components of the dielectric permittivity tensor near the "Minkowski domain wall". Similar to [17], continuous behavior of the tensor components in the thin transition layer near $z=0$ is assumed. Note that the metamaterial is uniaxial everywhere except $z=0$.

**Figure 3.** (a) General case of a "Minkowski domain wall" between two identical uniaxial hyperbolic metamaterials, which may support a surface wave. The optical axis of the hyperbolic metamaterial is oriented at angles $\alpha_1$ and $\alpha_2$ with respect to the wall in the left and right Minkowski half spaces, respectively. (b) Experimental geometry, which realizes a "Minkowski domain wall" supporting surface wave solutions.

**Figure 4**. (a-c) Experimental demonstration of world line behavior in an "expanding universe" using a plasmonic hyperbolic metamaterial: Optical (a) and AFM (b) images of the plasmonic hyperbolic metamaterial based on PMMA stripes on gold. The defect used as a plasmon source is shown by an arrow. (c) Plasmonic rays or "world lines" increase their spatial separation as a function of "timelike" radial coordinate. The point (or moment) $r=\tau=0$ corresponds to a toy "big bang". For the sake of clarity, light scattering by the edges of the PMMA pattern is partially blocked by semi-transparent triangles. (d-e) Experimental demonstration of the electromagnetic Minkowski domain

wall: (d) Optical image of the plasmonic metamaterial geometry imitating "colliding braneworlds" under white light illumination. (e) Same field of view illuminated with 488 nm laser light.

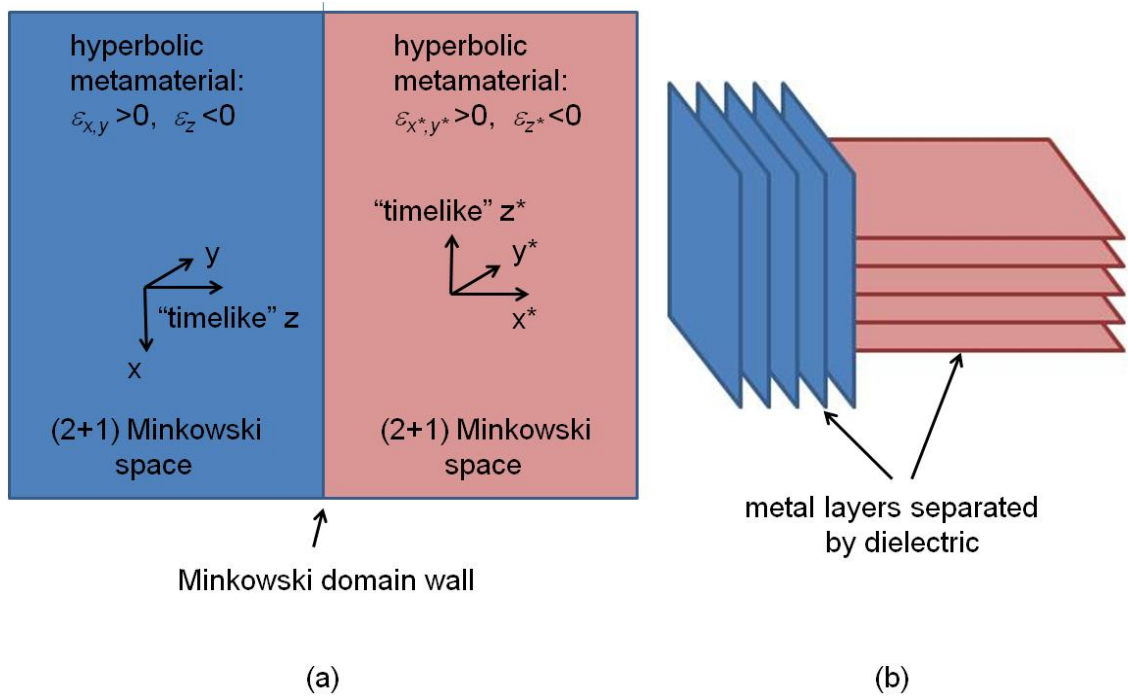

Fig.1

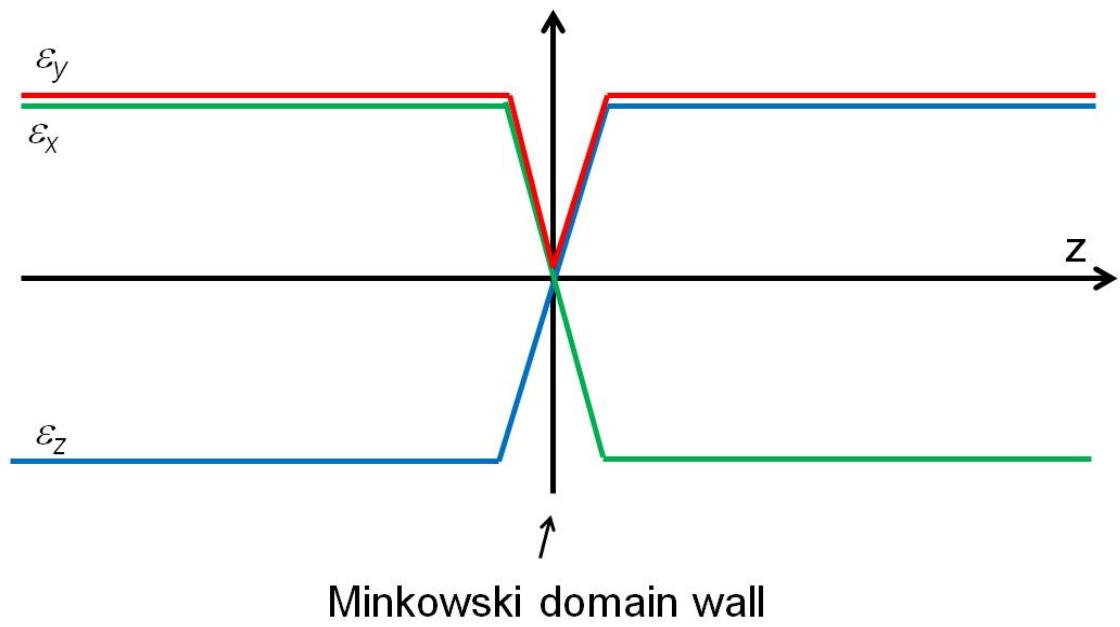

Fig.2

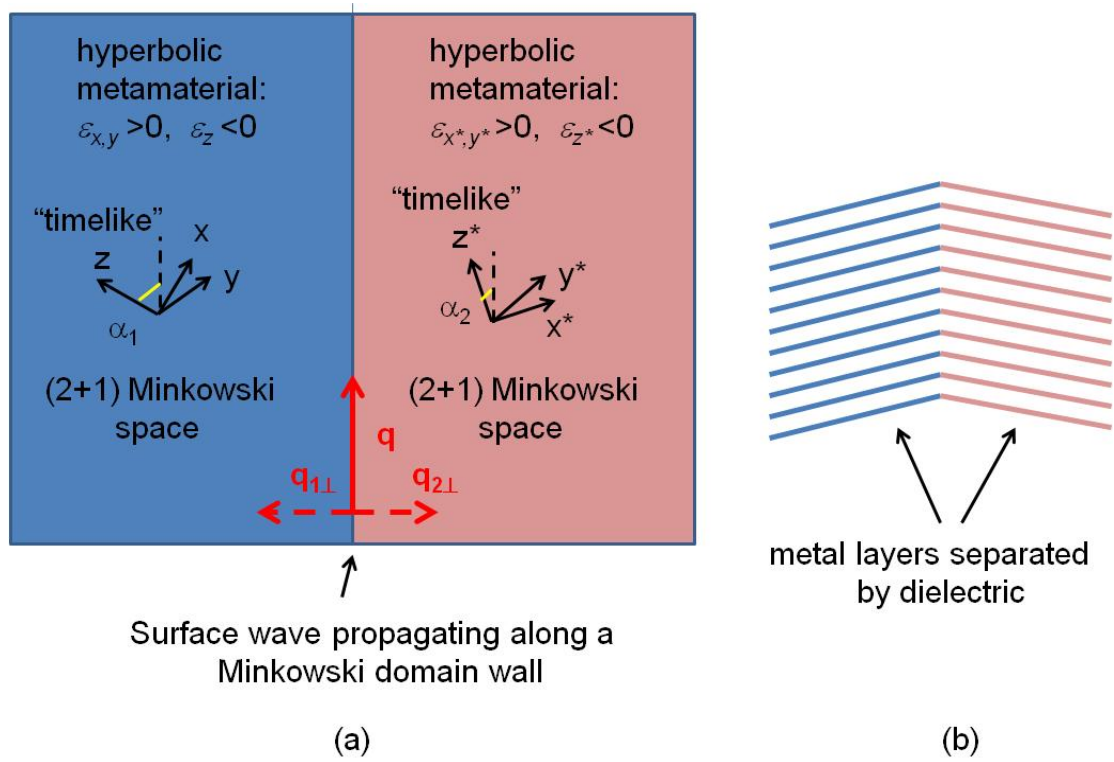

(a)                                                                 (b)

Fig.3

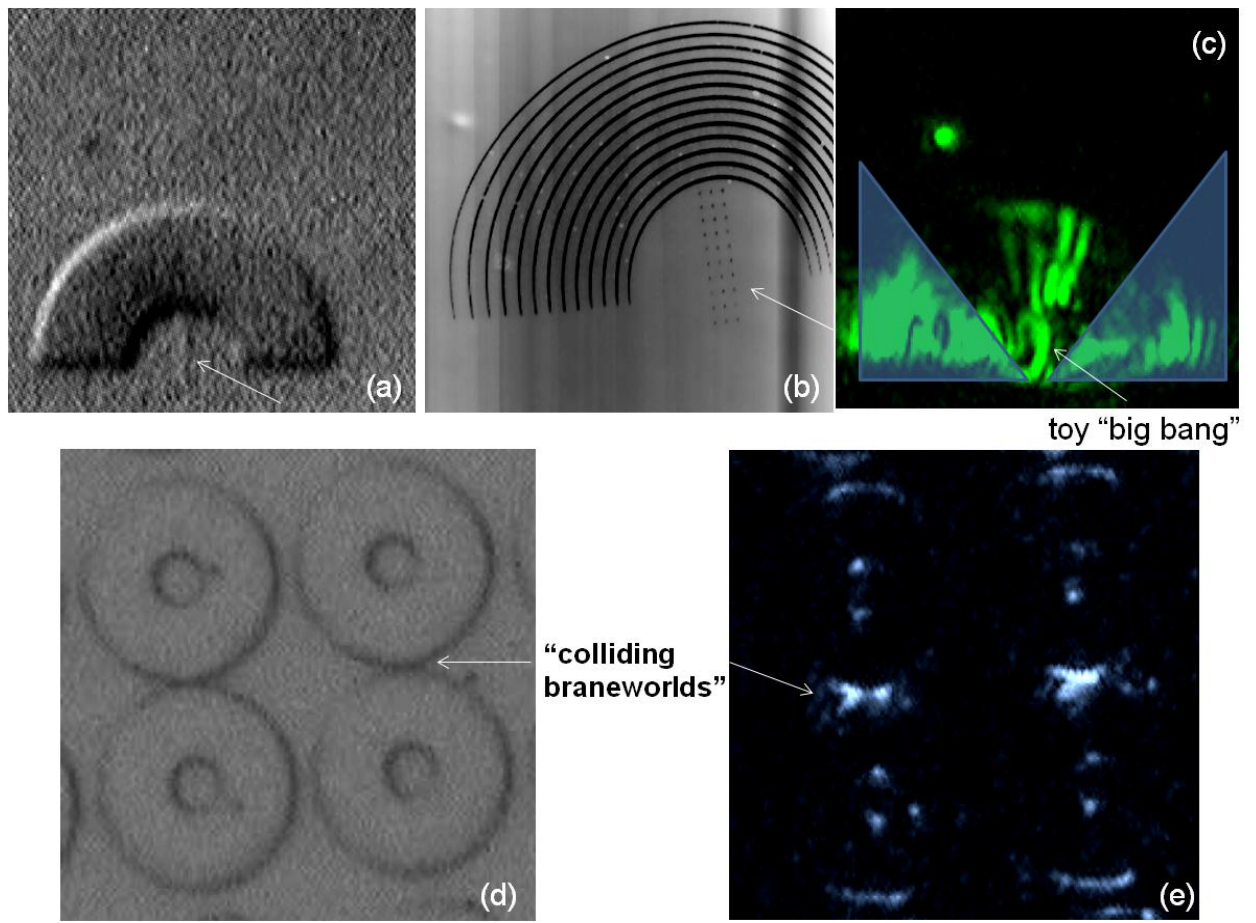

Fig.4